\documentclass[aps,pra,twocolumn,superscriptaddress,showpacs]{revtex4}
\usepackage{amssymb}
\usepackage{amsmath}
\usepackage{graphicx}
\usepackage{color}

\input{tcilatex}

\begin{document}

\title{Semi-quantum approach for fast atom diffraction: solving the rainbow
divergence}
\author{M.S. Gravielle\thanks{%
Author to whom correspondence should be addressed.\newline
Electronic address: msilvia@iafe.uba.ar}}
\affiliation{Instituto de Astronom\'{\i}a y F\'{\i}sica del Espacio. Consejo Nacional de
Investigaciones Cient\'{\i}ficas y T\'{e}cnicas. Casilla de Correo 67,
Sucursal 28, (C1428EGA) Buenos Aires, Argentina.}
\author{J.E. Miraglia}
\affiliation{Instituto de Astronom\'{\i}a y F\'{\i}sica del Espacio. Consejo Nacional de
Investigaciones Cient\'{\i}ficas y T\'{e}cnicas. Casilla de Correo 67,
Sucursal 28, (C1428EGA) Buenos Aires, Argentina.}
\date{\today }

\begin{abstract}
In this work we introduce a distorted wave method, based on the Initial
Value Representation (IVR) approach of the quantum evolution operator, in
order to improve the semiclassical description of rainbow effects in
diffraction patterns produced by grazing scattering of fast atoms from
crystal surfaces. The proposed theory, named Surface Initial Value
Representation (SIVR) approximation, is applied to He atoms colliding with a
LiF(001) surface along low indexed crystallographic channels. For this
collision system the SIVR approach provides a very good representation of
the quantum interference structures of experimental projectile
distributions, even in the angular region around classical rainbow angles
where common semiclassical methods diverge.
\end{abstract}

\maketitle

\section{Introduction}

The diffraction of fast atoms from crystal surfaces under grazing incidence
conditions has been the focus of extensive experimental and theoretical
research \cite%
{Schuller08,Manson08,Aigner08,Gravielle08,Bundaleski08,Schuller09PRB,Ruiz09,Diaz09,Khemliche09,Winter09,Schuller09PRL,Seifert10,Seifert12,Seifert13}
since its unexpected observation a few years ago \cite{Schuller07,Rousseau07}%
. From the theoretical point of view, different methods have been employed
to simulate experimental data of this phenomenon, now known as
grazing-incidence fast atom diffraction (GIFAD or FAD) \cite{Winter11}. They
range from full quantum treatments in terms of a wave packet propagation 
\cite{Rousseau07,Aigner08,Zugarramurdi12} to semiclassical approximations 
\cite{Schuller08,Gravielle08} based on the use of classical projectile
trajectories. \ Among these last theories we can mention the Surface Eikonal
(SE) approximation \cite{Gravielle08,SchullerGrav09}, which is a distorted
wave method that makes use of the eikonal wave function to represent the
elastic collision with the surface, while the motion of the fast projectile
is classically described by considering axially channeled trajectories for
different initial positions. The SE approach includes a clear description of
the main mechanisms of the FAD\ process, being simpler to evaluate than a
full quantum calculation \cite{Aigner08,Zugarramurdi12}. It has been applied
to investigate FAD patterns for different collision systems \cite%
{Gravielle09, SchullerGrav09,Gravielle11,Rubiano13}, showing a reasonable
agreement with the experiments in all the considered cases. \ 

In spite of the successful performance of the SE approach for the simulation
of FAD patterns, a weakness of the theory is its deficient description of
the rainbow effect, which affects the intensity of the outermost diffraction
maxima when these maxima are close to the classical rainbow angles \cite%
{Rubiano13}, i.e. the extreme deflection angles of the classical projectile
distribution . Such a deficiency, widely studied in atom-surface scattering 
\cite{MiretArtes12}, is a characteristic of \ the classical representation
of the collision dynamics, which introduces a singularity at rainbow angles
as a consequence of the presence of a point of accumulation of classical
trajectories (caustics), producing cusped rainbow peaks in the
angle-resolved scattering probability. In quantum mechanics, instead, these
sharp rainbow peaks are replaced by smooth maxima that display an
exponentially decaying behavior outside classical rainbow angles, just on
the dark side, i.e. in the region of classically forbidden transitions \cite%
{Berry72}. The goal of this article is to develop a semi-quantum
approximation for FAD, based on the Initial Value Representation (IVR)
method by Miller \cite{Miller70}, which can solve the drawback of the SE
model without losing the simple description of the interference process in
terms of classical scattering trajectories.

The IVR method represents a practical way of introducing quantum effects,
such as interferences and classical forbidden processes, in classical
dynamic simulations \cite{Miller01}. Taking as starting point the Feynman
path integral formulation of quantum mechanics, the basic idea of the IVR
method is to introduce the standard Van Vleck approximation \cite%
{VanVleck,Tannor} of the quantum evolution operator without considering any
additional assumption. That is, within the IVR model the full quantum \ time
evolution operator is replaced by the Van Vleck propagator in terms of
classical trajectories with different initial conditions, which is evaluated
numerically without using the common stationary phase approximation \cite%
{Miller01}. Precisely, this IVR strategy makes it possible to avoid the
classical rainbow divergence, incorporating an approximate description of
classically forbidden transitions in terms of real-valued trajectories \cite%
{Miller70}. The IVR solution has been successfully applied to different
branches, providing accurate transition probabilities for several atomic,
molecular and nuclear processes \cite%
{Miller70,Miller01,Sun97,Sun97b,Sun98,Skinner99,Dasso07}. In most of these
cases, IVR results are in excellent agreement with the corresponding full
quantum values.

In this paper we extend the IVR method to deal with FAD processes by using
the IVR\ time evolution operator in the frame of a time-dependent
distorted-wave formalism. The approach proposed here, named Surface-Initial
Value Representation (SIVR) approximation, is applied to evaluate FAD
patterns for He atoms grazing impinging on a LiF(001) surface. This
collision system will be used as a benchmark of the SIVR theory, comparing \
the results with available experimental data and with values derived within
the SE approach.

The article is organized as follows. The theoretical formalism is derived in
Sec. II. Results are presented and discussed in Sec. III, while in Sec. V we
outline our conclusions. Atomic units (a.u.) are used unless otherwise
stated.

\section{Theoretical model}

When an atomic projectile ($P$) \ grazingly impinges on a crystal surface ($%
S $) with an incidence energy $E$, the scattering state of the projectile, $%
\left\vert \Psi _{i}^{+}(t)\right\rangle $, satisfies the time-dependent Schr%
\"{o}dinger equation for the Hamiltonian 
\begin{equation}
H=-\frac{1}{2m_{P}}\nabla _{\vec{R}}^{2}+V_{SP}(\vec{R}),  \label{H}
\end{equation}%
where $\vec{R}\ $denotes the position of the center of mass of the incident
atom, $m_{P}$ is projectile mass, and $V_{SP}$ is the surface-projectile
interaction. The sign "$+$" in the scattering state indicates that it
satisfies outgoing asymptotic conditions for the elastic collision process,
verifying as initial condition that at $t=0$, when the projectile is far
from the surface, $\Psi _{i}^{+}$ tends to the state $\Phi _{i}$, where 
\begin{equation}
\Phi _{j}(\vec{R},t)=(2\pi )^{-3/2}\exp (i\vec{K}_{j}\cdot \vec{R}%
-iEt),\quad j=i(f)  \label{fi-i}
\end{equation}%
is the initial (final) unperturbed wave function, with $\vec{K}_{i}$ ($\vec{K%
}_{f}$) the initial (final) momentum and $%
E=K_{i}^{2}/(2m_{P})=K_{f}^{2}/(2m_{P})$.

In the Schr\"{o}dinger picture of quantum mechanics, the scattering state at
a given time $t$ can be formally expressed in terms of the evolution
operator $U(t)=\exp (-iH\ t)$ as 
\begin{equation}
\left\vert \Psi _{i}^{+}(t)\right\rangle =U(t)\ \left\vert \Phi
_{i}(0)\right\rangle  \label{Uoper}
\end{equation}%
for $t\geq 0$. A semiclassical expression of this equation can be obtained
by applying the IVR method, as summarized in Ref. \cite{Miller01}, to
represent the evolution operator $U(t)$. Within the IVR approach, the
scattering state of Eq. (\ref{Uoper}) becomes

\begin{eqnarray}
\left\vert \Psi _{i}^{+}(t)\right\rangle &\simeq &\left\vert \Psi
_{i}^{(IVR)+}(t)\right\rangle =(2\pi i)^{-3/2}\int d\overrightarrow{R}%
_{o}\int d\overrightarrow{K}_{o}  \notag \\
&&\times \ \left( J_{M}(t)\right) ^{1/2}\ \Phi _{i}(\overrightarrow{R}%
_{o},0)\exp (iS_{t})\left\vert \mathcal{\vec{R}}_{t}\right\rangle ,
\label{estado-ivr}
\end{eqnarray}%
where $\mathcal{\vec{R}}_{t}\equiv \mathcal{\vec{R}}_{t}(\overrightarrow{R}%
_{o},\overrightarrow{K}_{o})$ is the time-evolved position of the incident
atom at a given time $t$, which is obtained by considering a classical
trajectory with starting position and momentum $\overrightarrow{R}_{o}$ and $%
\overrightarrow{K}_{o}$, respectively. In Eq. (\ref{estado-ivr}) the
function $S_{t}\equiv S_{t}(\overrightarrow{R}_{o},\overrightarrow{K}_{o})$
denotes the classical action along the trajectory 
\begin{equation}
S_{t}=\int\limits_{0}^{t}dt^{\prime }\ \left[ \frac{\overrightarrow{\mathcal{%
P}}_{t^{\prime }}^{2}}{2m_{P}}-V_{SP}(\mathcal{\vec{R}}_{t^{\prime }})\right]
,  \label{St}
\end{equation}%
where $\overrightarrow{\mathcal{P}}_{t}\ $is the classical projectile
momentum at the time $t$, $\overrightarrow{\mathcal{P}}_{t}=m_{P}d\mathcal{%
\vec{R}}_{t}/dt$, while the function 
\begin{equation}
J_{M}(t)=\det \left[ \frac{\partial \mathcal{\vec{R}}_{t}(\overrightarrow{R}%
_{o},\overrightarrow{K}_{o})}{\partial \overrightarrow{K}_{o}}\right]
\label{J}
\end{equation}%
is a Jacobian factor (a determinant) evaluated along the classical
trajectory $\mathcal{\vec{R}}_{t}$, which is associated with the Maslov
phase. This Jacobian factor can be expressed as $J_{M}(t)=\left\vert
J_{M}(t)\right\vert \exp (i\nu _{t}\pi )$, where $\left\vert
J_{M}(t)\right\vert $ is the modulus of $J_{M}(t)$ and $\nu _{t}$ is an
integer number that accounts for the sign of $J_{M}(t)$ at a given time $t$.
In this way, $\nu _{t}$ represents a time-dependent Maslov index, satisfying
that every time that $J_{M}(t)$ changes its sign along the trajectory, $\nu
_{t}$ increases by 1.

In this work we use the IVR state of Eq. \ (\ref{estado-ivr}) to describe
the quantum scattering state within the framework of the time-dependent
distorted-wave formalism \cite{Dewangan94}. Hence, the distorted-wave
amplitude for the elastic transition from the initial state $\Phi _{i}$ to
the final state $\Phi _{f}$ can be expressed as 
\begin{equation}
A_{if}^{(SIVR)}=-i\int\limits_{0}^{+\infty }dt\left\langle \Phi _{f}\left(
t\right) \left\vert V_{SP}\right\vert \Psi _{i}^{(IVR)+}(t)\right\rangle .
\label{Aivr}
\end{equation}%
By replacing Eq. (\ref{estado-ivr})\ in Eq. (\ref{Aivr}) and explicitly
solving the integration on the spatial coordinate $\vec{R}$, which leads to
a Dirac delta function in the coordinate space, the SIVR transition
amplitude\ per unit of surface area reads 
\begin{equation}
A_{if}^{(SIVR)}=\frac{1}{\mathcal{S}}\int\limits_{\mathcal{S}}d%
\overrightarrow{R}_{os}\int d\overrightarrow{K}_{o}\ a_{if}^{(SIVR)}(%
\overrightarrow{R}_{o},\overrightarrow{K}_{o}),  \label{A-ivr}
\end{equation}%
where $\overrightarrow{R}_{o}=\overrightarrow{R}_{os}+Z_{o}\widehat{z}$ is
the starting position, at $t=0$, of the projectile trajectory, with $%
\overrightarrow{R}_{os}$ and $Z_{o}$ the components parallel and
perpendicular, respectively, to the surface plane, the $\hat{z}$ versor
oriented perpendicular to the surface, aiming towards the vacuum region, and 
$Z_{o}\rightarrow +\infty $. In Eq. (\ref{A-ivr}) the position $%
\overrightarrow{R}_{os}$ is integrated on a given area $\mathcal{S}$ of the
surface plane, the starting momentum $\overrightarrow{K}_{o}$ satisfies the
energy conservation, i.e. $\left\vert \overrightarrow{K}_{o}\right\vert
\equiv K_{0}=\sqrt{2m_{P}E}$, and 
\begin{eqnarray}
a_{if}^{(SIVR)}(\overrightarrow{R}_{o},\overrightarrow{K}_{o}) &=&\
-\int\limits_{0}^{+\infty }dt\ \ \frac{\left\vert J_{M}(t)\right\vert
^{1/2}e^{i\nu _{t}\pi /2}}{(2\pi i)^{9/2}}V_{SP}(\mathcal{\vec{R}}_{t}) 
\notag \\
&&\times \exp \left[ i\left( \varphi _{t}^{(SIVR)}-\overrightarrow{Q}\cdot 
\overrightarrow{R}_{o}\right) \right] ,\quad  \label{aif-ivr}
\end{eqnarray}%
is the SIVR transition amplitude associated with the classical path $%
\mathcal{\vec{R}}_{t}\equiv \mathcal{\vec{R}}_{t}(\overrightarrow{R}_{o},%
\overrightarrow{K}_{o})$, where $\overrightarrow{Q}=\vec{K}_{f}-\vec{K}_{i}$
is the projectile momentum transfer and $\varphi
_{t}^{(SIVR)}=Et+S_{t}-\Delta _{t}$ is the SIVR phase, with $\Delta _{t}=%
\vec{K}_{f}\cdot (\mathcal{\vec{R}}_{t}-\overrightarrow{R}_{o})$.

After some steps of algebra, the SIVR phase can be expressed as 
\begin{equation}
\varphi _{t}^{(SIVR)}=\int\limits_{0}^{t}dt^{\prime }\ \left[ \frac{1}{2m_{P}%
}\left( \vec{K}_{f}-\overrightarrow{\mathcal{P}}_{t^{\prime }}\right)
^{2}-V_{SP}(\mathcal{\vec{R}}_{t^{\prime }})\right] ,  \label{fitot}
\end{equation}%
which helps to reduce numerical uncertainties due to the fact that the
component of $\vec{K}_{i}$ parallel to the surface is much higher than the
perpendicular one. It is interesting to note that in Eq. (\ref{aif-ivr}),
the Jacobian factor $J_{M}(t)$ goes to zero as the Maslov index $\nu _{t}$
changes discontinuously, making the integrand be continuos at such a point 
\cite{Sun98}.

The SIVR differential probability, per unit of surface area, for elastic
scattering with final momentum $\vec{K}_{f}$ in the direction of the solid
angle $\Omega _{f}\equiv (\theta _{f},\varphi _{f})$ is obtained from Eq. (%
\ref{A-ivr}) as $dP/d\Omega _{f}=K_{f}^{2}\left\vert
A_{if}^{(SIVR)}\right\vert ^{2}$, where $\theta _{f}$ and $\varphi _{f}$ are
the final polar and azimuthal angles, respectively, with $\varphi _{f}$ \
measured with respect to the $\widehat{x}$ axis along the incidence
direction in the surface plane. A schematic depiction of the process and the
angular coordinates is displayed in Fig. 1.

\section{Results}

We apply the SIVR method to $^{4}$He atoms elastically scattered from a
LiF(001) surface under axial surface channeling conditions. This collision
system has been widely investigated with FAD \cite%
{Schuller08,Manson08,Aigner08,Gravielle08,Zugarramurdi12,SchullerGrav09,Gravielle09,Schuller10}
and will be considered as a benchmark for the theory.

The SIVR transition amplitude was obtained from Eq. (\ref{A-ivr}) \
employing the MonteCarlo technique to evaluate the $\vec{R}_{os}$ and $%
\overrightarrow{K}_{o}$ integrals. The integration on $\vec{R}_{os}$ was
done using random values obtained from a Gaussian distribution, while the
integral on \ $\overrightarrow{K}_{0}$ was solved making use of the change
of variables $\overrightarrow{K}_{0}=K_{0}(\cos \theta _{o}\cos \varphi
_{o},\cos \theta _{o}\sin \varphi _{o},-\sin \theta _{o})$, with $\theta
_{o} $ and $\varphi _{o}$ varying uniformly around the incidence direction,
in a range determined from the uncertainty principle. That is, the $\theta
_{o}$ and $\varphi _{o}$ variables were integrated in the angular ranges $%
\Delta \theta _{o}$ $\simeq \pm 10(K_{is}d_{z})^{-1}$ and $\Delta \varphi
_{o}\simeq \pm 5(K_{is}d_{y})^{-1}$, respectively, around the incidence
direction, where $K_{is}=K_{i}\cos \theta _{i}$ is the initial momentum
parallel to the surface, $\theta _{i}$ is the glancing incidence angle, and $%
d_{y}$ and $d_{z}$ are the lattice parameters in the directions $\widehat{y}$
and $\widehat{z}$, respectively, both of them perpendicular to the incidence
channel ($\widehat{x}$ axis). \ More than $4\times 10^{5}$ values of $\vec{R}%
_{os}$ and $\overrightarrow{K}_{o}$ were used in the calculation of $%
A_{if}^{(SIVR)}$ for each incidence condition, determined by the initial
momentum $\vec{K}_{i}$. It involved the sum of the $a_{if}^{(SIVR)}$%
amplitudes corresponding to different values of $\vec{R}_{os}$ and $%
\overrightarrow{K}_{o}$ that lead to the \textit{same} final momentum $\vec{K%
}_{f}$. This was done using a grid of $100\times 100$ points for the angles $%
\theta _{f}$ and $\varphi _{f}$ . Every transition amplitude $%
a_{if}^{(SIVR)} $ was evaluated numerically along the classical trajectory $%
\mathcal{\vec{R}}_{t}(\overrightarrow{R}_{o},\overrightarrow{K}_{o})$ from
Eq. (\ref{aif-ivr}). In such a calculation, the evaluation of the
determinant $J_{M}(t)$ represents the numerical bottleneck.

A key quantity \ to describe the experimental FAD patterns is the potential $%
V_{SP}$, which is here determined from a pairwise additive hypothesis by
adding individual contributions corresponding to the interaction of the
projectile with the different solid ions. Within this model, successfully
employed in FAD from insulator surfaces \cite{Gravielle09,Gravielle11}, the
surface-projectile potential takes into account the static and polarization
contributions. The static potential, derived by assuming that the electronic
densities of the particles remain frozen during the collision, was evaluated
as the sum of the electrostatic, kinetic and exchange potentials \cite%
{Gordon}. While in previous articles \cite{Gravielle09,Gravielle11} only
local electronic density contributions were considered, in this paper we
incorporate no local terms \ to evaluate the kinetic and exchange
potentials, as given by the Lee-Lee-Parr \cite{LLP} and Becke \cite{Becke}
models, respectively. In turn, the polarization potential, due to the
rearrangement of the projectile electron density induced by the presence of
target ions, was derived as in Ref. \cite{Gravielle11}. In addition, in the
calculation of the static and polarization contributions we have considered
a surface rumpling, with a displacement distance extracted from the
ab-initio calculation of Ref. \cite{SchullerGrav09}. Details of the
surface-potential calculation will be publish elsewhere \cite{Miragliatobe}.

\bigskip

As our main interest lies in analyzing the performance of the SIVR approach
to describe rainbow effects, first we \ focus on the mechanism of
supernumerary rainbows, which is associated with the SIVR amplitude derived
from Eq. (\ref{A-ivr}) by considering an area $\mathcal{S}$ equal to only 
\textit{one} reduced unit cell \cite{Schuller08,SchullerGrav09,Gravielle11}.
In Fig. 2 we compare SIVR projectile distributions for a reduced unit cell
with experimental data from Ref. \cite{Schuller08} for incidence along the $%
[100]$ channel with two different impact energies. The SIVR spectra, as a
function of the deflection angle $\Theta $ defined as $\Theta =\arctan
(\varphi _{f}\ /\theta _{f})$, present well defined peaks, which can be
identified as supernumerary rainbow maxima \cite{Schuller08}. The positions
and relative intensities of \ such peaks are in quite good agreement with
the experimental data, even the rainbow maximum which presents the highest
intensity. In contrast with previous semiclassical calculations \cite%
{Schuller08,Gravielle08,SchullerGrav09,Schuller12}, within the SIVR
approximation the rainbow peak is described as a smoothed maximum that takes
into account the decreasing intensity on the dark side of the classical
rainbow angle $\Theta _{rb}$ \cite{Berry72}. The angle $\Theta _{rb}$
corresponds to the largest deflection suffered for projectiles moving along
classical trajectories with initial momentum $\vec{K}_{i}$, so that
projectile paths ending with deflections $\Theta $ larger than $\Theta _{rb}$
are classically forbidden. Notice that \ despite the fact that no
convolution was introduced in the SIVR spectra of Fig. 2, the SIVR
probability displays a smooth behavior in the whole angular range, with a
gentle change of slope as a function of $\Theta $, in accord with the
experimental distribution. Theoretical spectra are expected to be symmetric
with respect to the incidence direction, which corresponds to the deflection
angle $\Theta =0$, while the experimental data are affected by experimental
uncertainties that partially break such a mirror symmetry.

Similar agreement between the SIVR and experimental distributions is also
observed for incidence along the $[110]$ channel, as shown in Fig. 3. For
this impact direction, the position and relative intensity of the rainbow
peak are properly reproduced by the SIVR approach. However, there is a
slight shift in the positions of the internal maxima, which is associated
with a failure of the surface-projectile interaction model for this channel.
As discussed in previous articles \cite{Aigner08,Gravielle09,Schuller10},
FAD patterns are extremely sensitive to the corrugation of the surface
potential across the incidence direction. Small differences in the potential
can strongly modify the positions of supernumerary rainbow maxima,
particularly, the internal ones, and this effect is more evident for
incidence along the [110] channel \cite{Gravielle09}.

With the aim of comparing the SIVR approach with previous semiclassical
theories \cite{Gravielle08,Gravielle09,SchullerGrav09}, in Fig. 4 we display
angular projectile distributions \ obtained with the SIVR and SE methods,
both of them including the supernumerary rainbow mechanism only, that is,
derived by integrating $\vec{R}_{os}$ on a reduced unit cell. In the case of
the semiclassical SE approximation, to study the role of the Maslov phase in
this new context, we have considered two versions: one incorporating the
Maslov phase \cite{SchullerGrav09} and the other without it \cite%
{Gravielle08,Gravielle09}. Within the SE approximation, the Maslov phase
represents a correction term $\phi _{M}^{(SE)}=\nu _{o}\pi /2$ that was
added to the phase of the scattering state in order to take into account the
phase change suffered by the wave as it passes through a focus, with $\nu
_{o}$ the Maslov index defined as in Ref. \cite{Avrin94}. From Fig. 4 we
observe that, like other semiclassical theories \cite%
{Schuller08,Winter11,Schuller12}, both versions of the SE approach produce
an abrupt increase of the probability at classical rainbow angles $\pm
\Theta _{rb}$, with null probability outside this angular range, on the dark
side of the rainbow angle. This deficiency is completely solved by the SIVR
method, which gives rise to smooth rainbow peaks with softened decaying
intensities for deflection angles larger than $\Theta _{rb}$. The most
important point to remark about the SIVR method is that the numerical
integration on the starting momentum $\overrightarrow{K}_{0}$, included in
Eq. (\ref{A-ivr}), regularizes the divergence of the transition amplitude
close to $\Theta _{rb}$, in such a way that forbidden trajectories as well
as the so called Airy behavior of the quantum transition amplitude are
automatically taken into account \cite{Miller01}.

Moreover, from Fig. 4 we found that the experimental positions of
supernumerary rainbow maxima are well described by the SE approach \textit{%
without} the Maslov correction term \cite{Gravielle08,Gravielle09}. But the
agreement deteriorates when the Maslov phase $\phi _{M}^{(SE)}$ is added to
the SE theory, turning the central minimum into a maximum, in contrast with
the experiment. On the contrary, the present SIVR approximation does
incorporates a similar Maslov phase $\phi _{M}(t)=\nu _{t}\pi /2$ as a
function of time along the classical trajectory. But this phase emerges
naturally, together with the factor $\left\vert J_{M}(t)\right\vert $, in
the derivation of Eq. (\ref{aif-ivr}), becoming in fact essential to obtain
proper projectile distributions within the SIVR method. Therefore, the
present results seem to indicate that\ the incorporation of $\phi
_{M}^{(SE)} $ in the SE approach is unbalanced, in a certain manner, and it
would be better to disregard it.

So far we have described the mechanism of supernumerary rainbows only, but
as it happens for most of the diffraction phenomena, FAD patterns have two
different origins: supernumerary rainbows and Bragg diffraction \cite%
{Schuller08}. Both mechanisms are included in the SIVR description and can
be analyzed separately, like in the SE approach \cite{Gravielle11}. In Eq. (%
\ref{A-ivr}) the integration region on the surface plane, $\mathcal{S}$, is
in principle determined by the size of the initial wave packet of incident
projectiles \cite{Joachain}. By considering this area as composed by $n$
identical reduced unit cells, each of them centered on a different site $%
\vec{X}_{sj}$ of the crystal surface, we can express the corresponding SIVR
transition amplitude as 
\begin{equation}
A_{if,n}^{(SIVR)}=A_{if,1}^{(SIVR)}\ S_{n}(\vec{Q}_{s}),  \label{Aivr-tot}
\end{equation}%
where $A_{if,1}^{(SIVR)}$ is derived from Eq. (\ref{A-ivr}) by evaluating
the $\vec{R}_{os}$-integral over \textit{one} reduced unit cell, while the
function 
\begin{equation}
S_{n}(\vec{Q}_{s})\ =\frac{1}{n}\sum\limits_{j=1}^{n}\exp \left[ -i\vec{Q}%
_{s}.\vec{X}_{sj}\right]  \label{Sq}
\end{equation}%
takes into account the crystallographic structure of the surface, with $\vec{%
Q}_{s}$\ the component of $\vec{Q}$ parallel to the surface plane. Each
factor in Eq. (\ref{Aivr-tot}) describes a different mechanism. The factor $%
A_{if,1}^{(SIVR)}$ is related to supernumerary rainbows and carries
information on the shape of the interaction potential across the incidence
channel, while the factor $S_{n}(\vec{Q}_{s})$ is associated with the Bragg
diffraction and provides information on the spacing between surface atoms
only. As the component of the momentum transfer along the incidence channel
is negligible, we can approximate $S_{n}(\vec{Q}_{s})\ \approx S_{n}(Q_{%
\mathrm{tr}})$, where $Q_{\mathrm{tr}}=K_{f}\cos \theta _{f}\sin \varphi
_{f} $ is the component of the transferred momentum transversal to the
incidence channel on the surface plane. For scattering along the $[110]$
channel this function reads%
\begin{equation}
S_{n}^{[110]}(Q_{\mathrm{tr}})=\frac{\sin (n_{\mathrm{tr}}\beta )}{n_{%
\mathrm{tr}}\sin \beta },  \label{S110}
\end{equation}%
while for incidence along the $[100]$ channel it reads 
\begin{equation}
S_{n}^{[100]}(Q_{\mathrm{tr}})=(n_{\mathrm{tr}}^{2}+1)^{-1}\left[ n_{\mathrm{%
tr}}^{2}\frac{\sin (n_{\mathrm{tr}}\beta )}{n_{\mathrm{tr}}\sin \beta }+%
\frac{\cos (n_{\mathrm{tr}}\beta )}{\cos \beta }\right] ,  \label{S100}
\end{equation}%
where $n_{\mathrm{tr}}$\ is the number of reduced unit cells along the
transverse direction (fixed as an odd number) and $\beta =Q_{\mathrm{tr}}\
d/2$, with $d$ the spatial lattice periodicity of the channel. Hence, in
both directions $S_{n}(Q_{\mathrm{tr}})$ gives rise to Bragg maxima placed at%
\begin{equation}
Q_{\mathrm{tr}}d=m2\pi ,  \label{Bragg}
\end{equation}%
with $m$ \ an integer number. The width of these diffraction peaks is
affected by the number of reduced unit cells reached by the incident wave
packet, i.e. the larger $n_{\mathrm{tr}}$\ is, the narrower the Bragg peaks
are.

To visualize the above behavior, in Fig. 5 \ we display the SIVR
distribution obtained from Eq. (\ref{A-ivr}) by integrating $\vec{R}_{os}$
on an area $\mathcal{S}$ formed by \textit{three} reduced unit cells. In
this case, the SIVR spectrum presents Bragg maxima as superimposed
structures to the supernumerary contribution. Resolved Bragg peaks can be
observed in experimental projectile distributions for low values of the
perpendicular energy $E_{\perp }=E\sin ^{2}\theta _{i}$, associated with the
motion normal to the surface plane \cite{SchullerGrav09}. But for higher
perpendicular energies, like the ones considered in Figs. 2 and 3, discrete
Bragg peaks originated from the interference of trajectories from different
reduced unit cells are not present in the experimental distributions due to
the limits in spatial resolution of the detector \cite{Gravielle11}.
Therefore, only supernumerary rainbow contributions are visible in the
experimental spectra of such figures.

\section{Conclusions}

We have developed a semi-quantum approximation based on the IVR method of
Miller%
\'{}%
s \cite{Miller70} to deal with FAD from crystal surfaces. The proposed
approach - the SIVR approximation - solves the rainbow singularities
originated by the classical description of the projectile dynamics,
preserving a simple semi-quantum picture of the main mechanisms of the
process. In order to test the reliability of the SIVR method, we have
applied it to keV He atoms colliding under grazing incidence with a LiF(001)
surface, for which there are available experimental data. The surface
potential was derived from a pairwise additive model, including non local
kinetics and exchange contributions, polarization and rumpling. From the
comparison of calculated angular spectra with experimental projectile
distributions for two different low-indexed crystallographic directions of
the LiF surface we conclude that the SIVR approach provides a very good
representation of the FAD patterns in the whole angular range, without
requiring the use of convolutions to smooth the theoretical curves.
Therefore, the SIVR method might be considered as an attractive alternative
to quantum wave packet propagations, offering a realistic description of FAD
patterns, even around classical rainbow angles.

We also found that the use of the Maslov correction term in the SE
approximation might be inadequate, while in the SIVR approximation the
Maslov phase emerges as a function of the projectile position along the
classical trajectory, playing an essential role.

\begin{acknowledgments}
M.S.G. is kindly grateful to Marcos Saraceno for his helpful suggestion.
M.S.G and J.E.M acknowledge financial support from CONICET, UBA, and ANPCyT
of Argentina.
\end{acknowledgments}

\begin{figure}[tbp]
\includegraphics[width=0.4\textwidth]{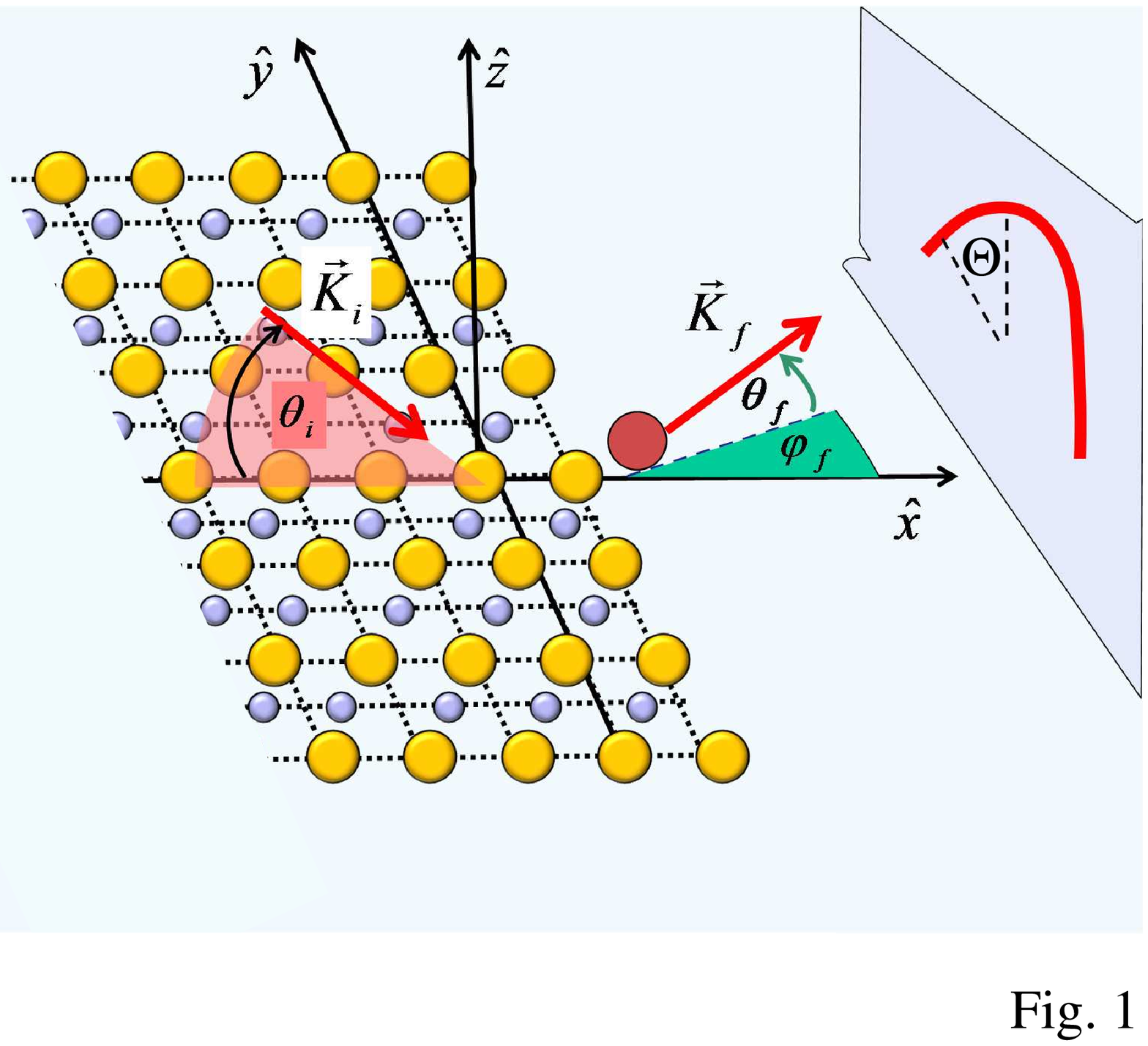}
\caption{(Color online) Sketch of the angular coordinates for the FAD
process. }
\label{Fig1}
\end{figure}

\begin{figure}[tbp]
\includegraphics[width=0.4\textwidth]{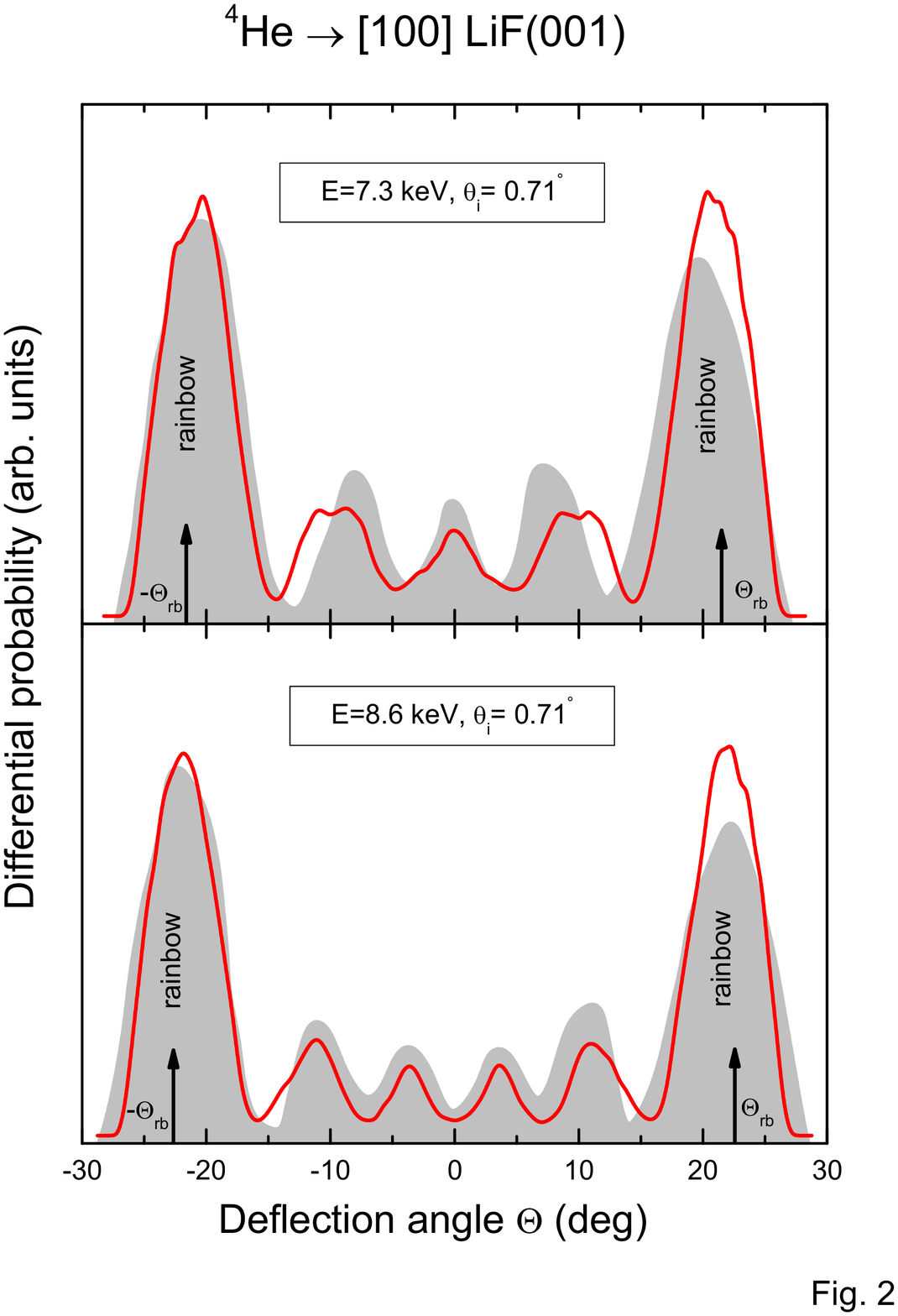}
\caption{(Color online) Angular projectile distribution, as a function of
the deflection angle $\Theta $, for $^{4}$He atoms scattered from LiF(001)
along the $[100]$ direction with the glancing incidence angle $\protect%
\theta _{i}=0.71$ deg. Two different impact energies are considered: (a) $E$ 
$=7.3$ keV; (b) $E$ $=8.6$ keV. Solid red line, SIVR results for the
supernumerary rainbow mechanism, derived by integrating the starting
position $\vec{R}_{os}$ over a reduced unit cell; shadow gray line,
experimental data from Ref. \protect\cite{Schuller08}. Vertical arrows,
positions of the classical rainbow angles $\pm \Theta _{rb}$.}
\label{Fig2}
\end{figure}

\begin{figure}[tbp]
\includegraphics[width=0.4\textwidth]{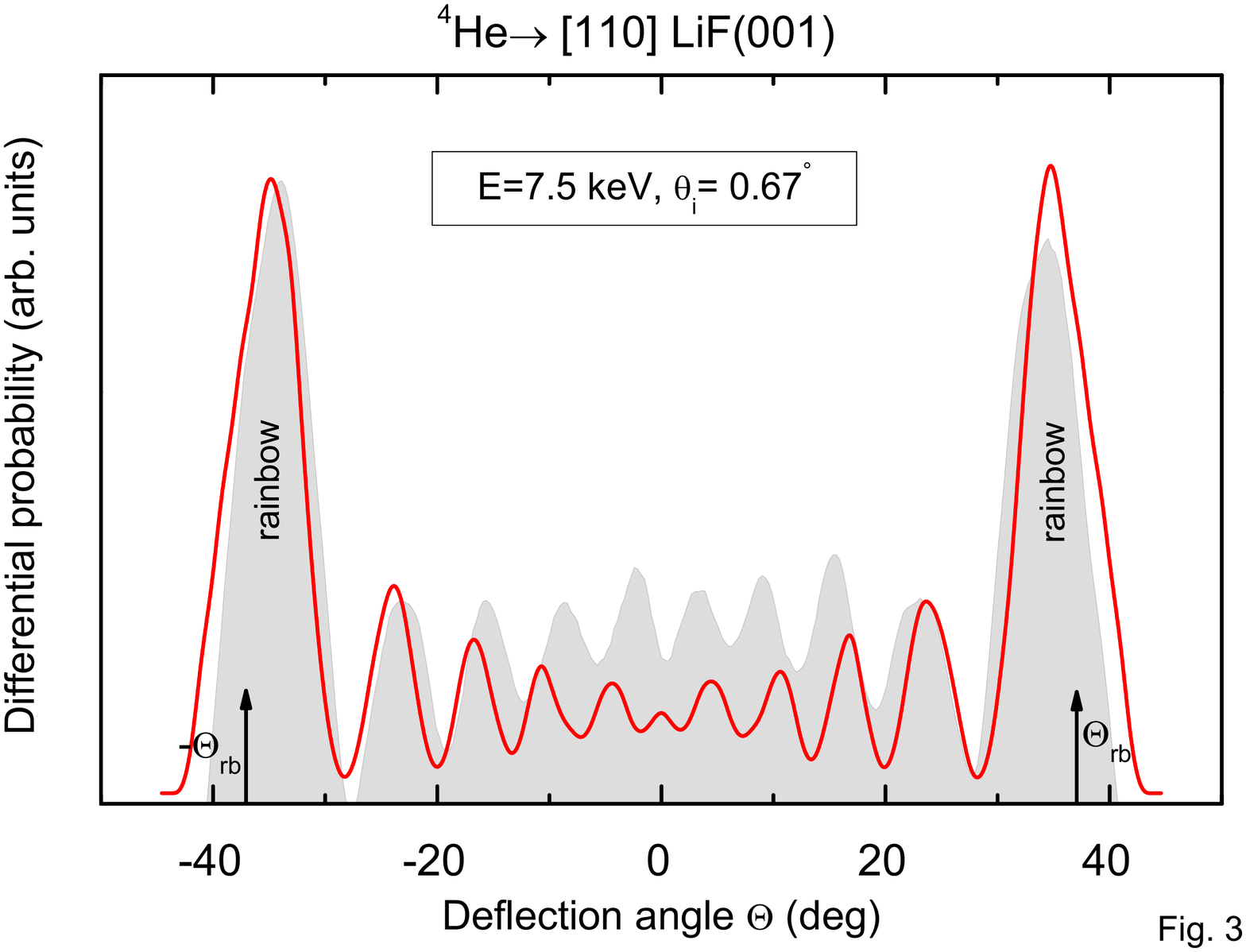}
\caption{(Color online) Similar to Fig. 2 for $^{4}$He atoms scattered from
LiF(001) along the $[110]$ channel. The incidence energy and angle are $E$ $%
=7.5$ keV and $\ \protect\theta _{i}=0.67$ deg., respectively. The
experimental data were extracted from Ref. \protect\cite{SchullerGrav09}.}
\label{Fig3}
\end{figure}

\begin{figure}[tbp]
\includegraphics[width=0.4\textwidth]{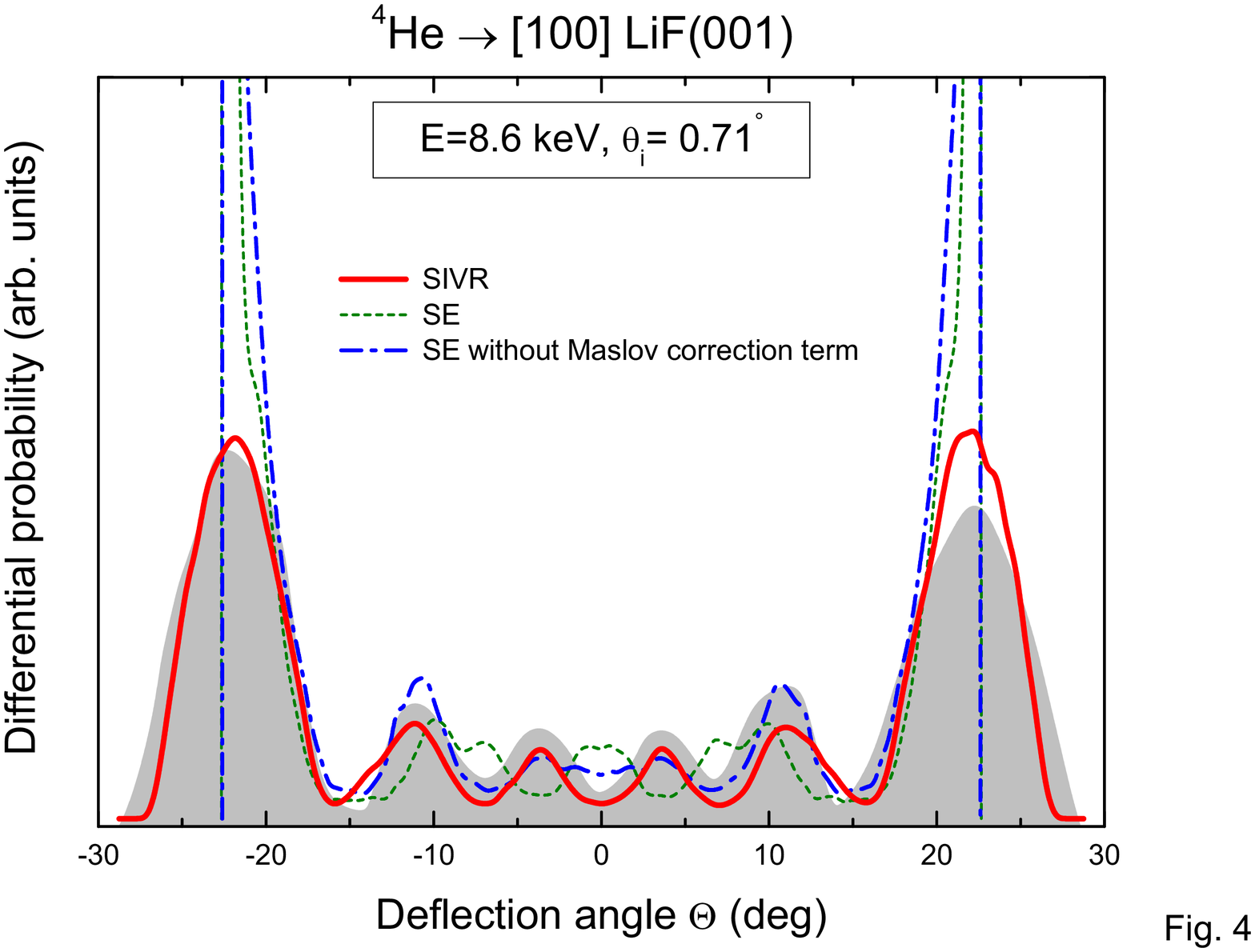}
\caption{(Color online) Similar to Fig. 2 (b) considering different
theoretical descriptions of the supernumerary rainbow mechanism. Solid red
line, SIVR approximation; dashed green line, SE approach; dash-dotted blue
line, SE approach without including the Maslov correction term, as explained
in the text.\ All the theories evaluated integrating the starting position $%
\vec{R}_{os}$ over a reduced unit cell.}
\label{Fig4}
\end{figure}

\begin{figure}[tbp]
\includegraphics[width=0.4\textwidth]{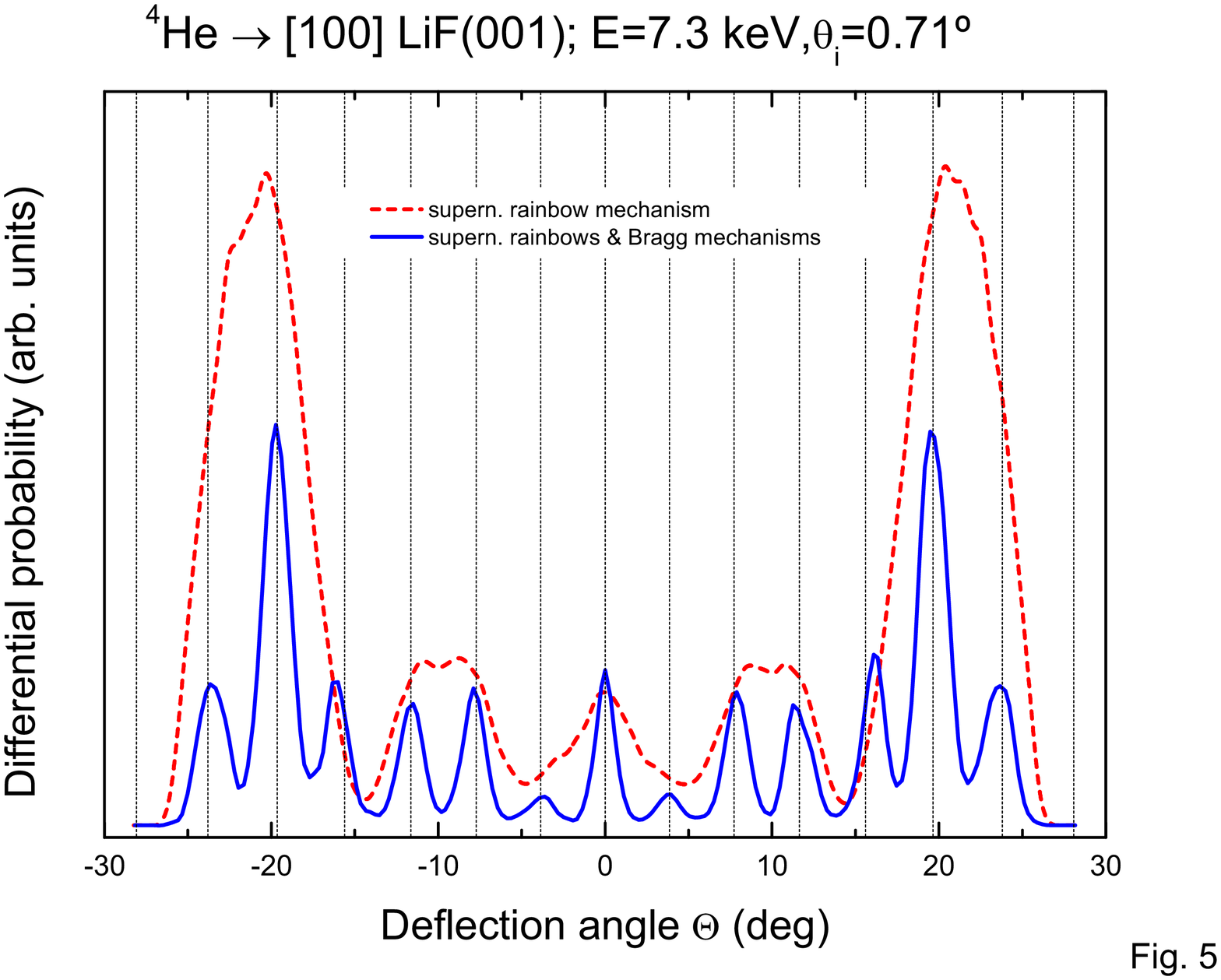}
\caption{(Color online) Similar to Fig. 2 (a) comparing the contributions of
the different mechanisms. Dashed red line, SIVR results derived from Eq. (%
\protect\ref{A-ivr}) by integrating the starting position $\protect%
\overrightarrow{R}_{os}$ over a reduced unit cell (supernumerary rainbow
contribution); solid blue line, similar by using an extended integration
area, as explained in the text. Dotted vertical lines, theoretical peak
positions based on the Bragg condition (Eq. (\protect\ref{Bragg})).}
\label{Fig5}
\end{figure}

\end{document}